\documentclass [prl,twocolumn,superscriptaddress,showpacs]{revtex4}
\usepackage {graphicx}

\begin{document}

% You may use Title,Subject,Author,Manager,Company,Operator,
% Category,Comment,Hlinkbase document properties here

\title{Differential Photoelectron Holography:\\
A New Approach for Three-Dimensional Atomic Imaging}
\begin{abstract}
We propose differential holography as a method to overcome the long-standing forward-scattering problem in 
photoelectron holography and related techniques for the three-dimensional imaging of atoms. 
Atomic images reconstructed from experimental and theoretical Cu 3p holograms from Cu(001) demonstrate that 
this method suppresses strong forward-scattering effects so as to yield more accurate three-dimensional images 
of side- and back-scattering atoms.
\end{abstract}
\author{S. Omori}
\affiliation{Institute of Industrial Science, University of Tokyo, Tokyo 153-8505, Japan}
\affiliation{Materials Sciences Division, Lawrence Berkeley National Laboratory,
Berkeley, California 94720}
\author{Y. Nihei}
\affiliation{Institute of Industrial Science, University of Tokyo, Tokyo 153-8505, Japan}
\author{E. Rotenberg}
\affiliation{Advanced Light Source, Lawrence Berkeley National Laboratory,
Berkeley, California 94720}
\author{J. D. Denlinger}
\affiliation{Advanced Light Source, Lawrence Berkeley National Laboratory,
Berkeley, California 94720}
\author{S. Marchesini}
\affiliation{Materials Sciences Division, Lawrence Berkeley National Laboratory,
Berkeley, California 94720}
\author{S. D. Kevan}
\affiliation{Department of Physics, University of Oregon,
Eugene, Oregon 97403}
\author{B. P. Tonner}
\affiliation{Department of Physics, University of Central Florida,
Orlando, Florida 32816}
\author{M. A. Van Hove}
\affiliation{Advanced Light Source, Lawrence Berkeley National Laboratory,
Berkeley, California 94720}
\affiliation{Materials Sciences Division, Lawrence Berkeley National Laboratory,
Berkeley, California 94720}
\affiliation{Department of Physics, University of California,
Davis, California 95616}
\author{C. S. Fadley}
\affiliation{Materials Sciences Division, Lawrence Berkeley National Laboratory,
Berkeley, California 94720}
\affiliation{Department of Physics, University of California,
Davis, California 95616}

\pacs{61.14.-x, 42.40.-i}

\date{\today}
\maketitle

Holography \cite{Gabor:1948} is a method of recording both the amplitudes 
and phases of waves scattered by an object illuminated with coherent 
radiation, and using this information to directly construct a 
three-dimensional image of the object. Sz\"{o}ke \cite{Szoke:1986} first 
suggested that coherent outgoing waves from atomically-localized sources of 
photoelectrons, fluorescent x-rays, and $\gamma $-rays could be used to 
achieve atomic-scale holography. This idea was initially demonstrated 
theoretically for the case of photoelectrons by Barton 
\cite{Barton:1988}, and then extended into a multi-energy format by 
Barton and Terminello and by Tong and co-workers \cite{Barton:1991}. By 
now several experimental approaches to such atomic-resolution holography 
have been demonstrated, including photoelectrons 
\cite{Tonner:1991,Terminello:1993,Tong:1995,Len:1999}, 
Auger electrons \cite{Saldin:1993}, Kikuchi electrons \cite{Wei:1994}, 
diffuse-scattered low-energy electrons \cite{Saldin:1998}, fluorescent 
x-rays in either a direct mode \cite{Tegze:1994} or a multi-energy 
inverse mode \cite{Gog:1996}, $\gamma $-rays \cite{Korecki:1997}, and 
bremsstrahlung x-rays \cite{Bompadre:1999}.

Among these methods, photoelectron holography (PH) has the advantages of 
being capable of studying the local atomic structure around each type of 
emitter without requiring long-range order and of distinguishing emitters 
through core-level binding-energy shifts \cite{Len:1999}. 
Photoelectron holograms also show strong modulations of up to $\pm $50{\%}, 
so such effects are easily measurable. However, PH can suffer from serious 
image aberrations due to the strength of electron scattering. The atomic 
scattering factor $f$ is a highly anisotropic function of scattering angle, and 
can depend strongly on electron kinetic energy $E_{k}$. In particular, as 
$E_{k}$ increases above a few hundred eV, $f$ becomes more and more significant 
in the forward direction, resulting in a strong forward-scattering (FS) peak 
\cite{Fadley:1993} that can induce image aberrations. Beyond this, PH 
also can suffer from multiple-scattering (MS) effects due to the scattering 
strength.

Various reconstruction algorithms and measurement methods 
\cite{Barton:1991,Tonner:1991,Tong:1995,Greber:1996} 
have been proposed to correct for the anisotropic $f$ and MS effects, some of 
which can be summarized via
\begin{equation}
\label{eq1}
U\left( {\rm {\bf r}} \right) = \left| {\;\int {\;W\chi \left( {\rm {\bf k}} 
\right)\;\exp \left[ { - ikr + i{\rm {\bf k}} \cdot {\rm {\bf r}}} 
\right]d^3{\rm {\bf k}}\;} } \right|^2
\end{equation}
\noindent
where $U$ is the image intensity at position \textbf{r}, \textit{$\chi $} is the normalized 3D 
hologram, and the function or operator $W$ permits describing the difference 
between algorithms, with $W$=1 in the original multi-energy formulations 
\cite{Barton:1991}. One alternative algorithm \cite{Tonner:1991} 
sets $W = f^{ - 1}\left( {k,\theta _{{\rm {\bf r}}}^{{\rm {\bf k}}} } 
\right)$ so as to divide out the anisotropic $f$, where $\theta _{{\rm {\bf 
r}}}^{{\rm {\bf k}}} $ is the angle between \textbf{r} and \textbf{k}. In 
another algorithm \cite{Tong:1995} based on the more ideal electron 
back scattering (BS), a window function for $W$ that limits the integral in Eq. 
(\ref{eq1}) to be in a small cone of ${\rm {\bf \hat {k}}}$ around $ - {\rm {\bf 
r}}$ is chosen to emphasize the imaging of BS atoms. Although successful in 
several applications \cite{Tong:1995,Luh:1998}, it is 
difficult to apply this small-cone method to many systems where the imaging 
of FS or even side-scattering (SS) atoms is important, such as epitaxial 
films and buried interfaces. In fact, imaging of ``bulk'' atoms surrounded by 
FS and BS atoms via PH has proven to be especially difficult [cf. Figs. 7-9 
in ref. \cite{Len:1999}], with most successful applications being to 
emitters in the first few layers near a surface.

To overcome such FS effects, we propose in this Letter ``differential 
holography''. By simply replacing $\chi $ in Eq. (\ref{eq1}) by its $k$-derivative 
(i.e. $W = \partial \mathord{\left/ {\vphantom {\partial {\partial k}}} 
\right. \kern-\nulldelimiterspace} {\partial k})$ or more conveniently by a 
numerical difference between two $\chi $'s at different energies ($\delta 
\chi = \chi \mbox{(}k + \delta k\mbox{)} - \chi \mbox{(}k\mbox{)})$, FS 
effects can be greatly suppressed. We have applied this method to 
multi-energy holograms for Cu 3p emission from Cu(001), and show that this 
provides images that are improved over prior work in several respects.

To avoid confusion with other methods in PH, we also note that 
``derivative'' PH has been proposed and used successfully by Chiang and 
co-workers \cite{Luh:1998}. However, the purpose here is to 
eliminate uncertainties in $I$ due to the variation of experimental 
conditions by first taking logarithmic derivatives ${\mbox{[}\partial I / 
\partial k\mbox{]}} \mathord{\left/ {\vphantom {{\mbox{[}\partial I / 
\partial k\mbox{]}} I}} \right. \kern-\nulldelimiterspace} I$ that are then 
reintegrated into ``self-normalized'' intensities; thus, it is still finally 
$\chi$ that is used in Eq.(\ref{eq1}).

\begin{figure}[htbp]
\centerline{\includegraphics[width=3.5in]{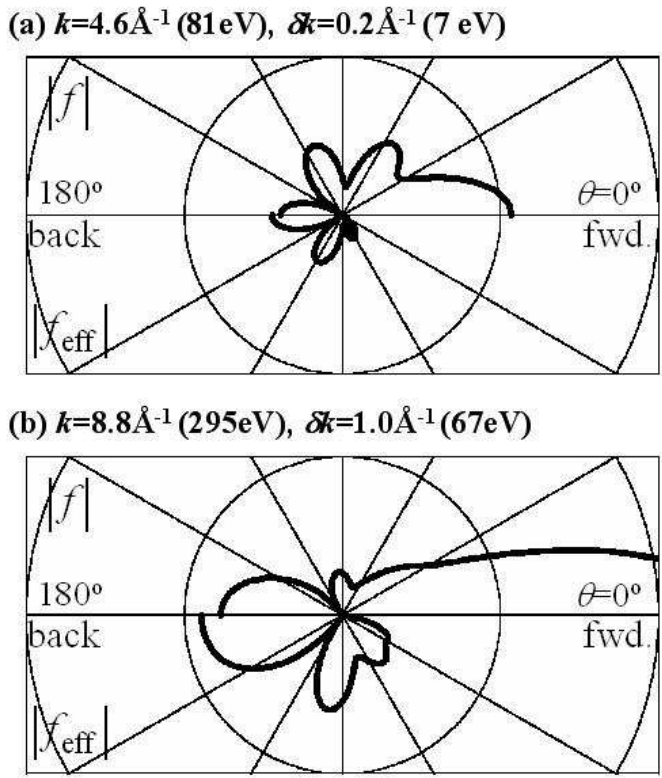}}
\caption{Comparison of the usual scattering amplitude $\left| f \right|$ and the 
effective scattering amplitude of differential holography $\left| 
{f_{\mbox{eff}} } \right|$, calculated for Cu-Cu nearest neighbors (r=2.56 
{\AA}) as a function of scattering angle $\theta _{{\rm {\bf r}}}^{{\rm {\bf 
k}}} $ for two different sets of $k$ and $\delta k$ in taking the 
differential of $\chi $: (a) $k$ = 4.6 {\AA}$^{ - 1}$ (81 eV), $\delta k$ = 
0.2 {\AA}$^{ - 1}$ (7 eV) and (b) $k$ = 8.8 {\AA}$^{ - 1}$ (295 eV), $\delta 
k$ = 1.0 {\AA}$^{ - 1}$ (67 eV). The final strong forward-scattering data 
points of $\left| f \right|$ at the right of panel (b) are truncated.}
\label{fig1}
\end{figure}

The principle of differential photoelectron holography (DPH) is as follows. 
We consider the single-scattering expression of $\chi $ for an 
emitter-scatterer pair spaced by a vector ${\rm {\bf r}}$ 
\cite{Fadley:1993}:
\begin{eqnarray}
\label{eq2}
 \chi \left( {\rm {\bf k}} \right) =&\frac{I - I_{0} }{I_{0} }\nonumber \\ 
 \approx&\frac{2\left| {f\left( {k,\theta _{{\rm {\bf r}}}^{{\rm {\bf k}}} } 
\right)} \right|}{r}\mbox{cos}\left[ {kr\left( {1 - \mbox{cos}\theta _{{\rm 
{\bf r}}}^{{\rm {\bf k}}} } \right) + \varphi \left( {k,\theta _{{\rm {\bf 
r}}}^{{\rm {\bf k}}} } \right)} \right],
\end{eqnarray}

\noindent
where $I_{0} $ is the intensity that would be observed without atomic 
scattering, and $\varphi $ is the scattering phase. If $\delta k$ is 
sufficiently small so that ${\delta \left| f \right|} \mathord{\left/ 
{\vphantom {{\delta \left| f \right|} {\left| f \right|}}} \right. 
\kern-\nulldelimiterspace} {\left| f \right|} \ll 1$, where $\delta \left| f 
\right|$ is the change in $\left| f \right|$, the difference of two 
holograms at $k_{\pm } = k\pm {\delta k} \mathord{\left/ {\vphantom {{\delta 
k} 2}} \right. \kern-\nulldelimiterspace} 2$ can be written in a similar 
form to Eq. (\ref{eq2}) as:

\begin{eqnarray}
\label{eq3}
 \delta \chi \left( {\rm {\bf k}} \right) =& \chi \left( {k_{ + } {\rm {\bf 
\hat {k}}}} \right) - \chi \left( {k_{ - } {\rm {\bf \hat {k}}}} \right)\nonumber \\ 
 \approx& - \frac{2\left| {f_{\mbox{eff}} } \right|}{r}\mbox{sin}\left[ 
{kr\left( {1 - \mbox{cos}\theta _{{\rm {\bf r}}}^{{\rm {\bf k}}} } \right) + 
\bar {\varphi }\left( {k,\theta _{{\rm {\bf r}}}^{{\rm {\bf k}}} } \right)} 
\right], 
\end{eqnarray}

\noindent
where direction $\hat {k}$ is defined by angles $\theta $ and $\phi $, the 
``effective'' scattering amplitude is defined as $\left| {f_{\mbox{eff}} } 
\right| = 2\left| f \right|\mbox{sin}\left[ {{\delta kr\left( {1 - 
\mbox{cos}\theta _{{\rm {\bf r}}}^{{\rm {\bf k}}} } \right)} \mathord{\left/ 
{\vphantom {{\delta kr\left( {1 - \mbox{cos}\theta _{{\rm {\bf r}}}^{{\rm 
{\bf k}}} } \right)} 2}} \right. \kern-\nulldelimiterspace} 2 + {\delta 
\varphi } \mathord{\left/ {\vphantom {{\delta \varphi } 2}} \right. 
\kern-\nulldelimiterspace} 2} \right]$, and $\bar {\varphi }$ is the average 
of $\varphi $'s at $k_{\pm } $. In the FS region where $\theta _{{\rm {\bf 
r}}}^{{\rm {\bf k}}} \to 0$, $\left| {f_{\mbox{eff}} } \right|$ is thus very 
small, approaching zero in the limit of $\delta \varphi \to 0$. If \textit{$\delta $k} is also 
small, $\left| {f_{\mbox{eff}} } \right|$ is proportional to $r$; thus, DPH not 
only suppresses the FS effects, but also enhances the imaging of distant 
atoms. In Fig. \ref{fig1}, $\left| f \right|$ and 
$\left| {f_{\mbox{eff}} } \right|$ are plotted as a function of $\theta 
_{{\rm {\bf r}}}^{{\rm {\bf k}}} $ for Cu-Cu nearest neighbors ($r$=2.56 
{\AA}). For $k$=4.6 {\AA}$^{ - 1}$ and \textit{$\delta $k}=0.2 {\AA}$^{ - 1}$, $\left| 
{f_{\mbox{eff}} } \right|$ is significant only in the region of $\theta 
_{{\rm {\bf r}}}^{{\rm {\bf k}}} > \sim 90^o$. Therefore, the imaging of SS 
and BS atoms is expected, while it will be difficult for this case to image 
FS atoms. On the other hand, for $k$=8.8 {\AA}$^{ - 1}$ and a larger fractional 
\textit{$\delta $k}=1.0 {\AA}$^{ - 1}$, $\left| {f_{\mbox{eff}} } \right|$ is significant not 
only in the BS region but also in the range of $\theta _{{\rm {\bf 
r}}}^{{\rm {\bf k}}}  \sim $30$^{o}$-90$^{o}$. Since near-neighbor FS 
diffraction fringes extend out beyond 30$^{o}$ 
\cite{Fadley:1993,Omori:1999}, we might expect the latter 
choice to also permit imaging FS atoms. In this way, the relative 
sensitivity of DPH to SS and FS atoms can be ``tuned'' by selecting the 
range and step width of $k$ scans. Finally, we note that the suppression of MS 
effects by means of a transform over a volume in \textbf{k} space is well 
known in normal multi-energy PH \cite{Barton:1991} and this 
suppression will be equally present in DPH. If anything, the inherent 
elimination of strong FS effects in DPH should lead to even better MS 
suppression.

To demonstrate DPH experimentally, photoelectron holograms from Cu(001) were 
measured at beamline 7.0 of the Advanced Light Source at the Lawrence 
Berkeley National Laboratory. Photoelectron spectra for Cu 3p emission were 
collected at 25 energies over $k$=4.5-9.3 {\AA}$^{ - 1}$ ($E_{k}$=77-330 eV) 
with a constant step of \textit{$\delta $k}=0.2 {\AA}$^{ - 1}$ (\textit{$\delta $E}$_{k}$ =7-14 eV), along 65 
different directions over a symmetry-reduced 1/8 of the total solid angle 
above the specimen, and with a polar angle range from \textit{$\theta $}=0$^{o}$ (surface 
normal) to 70$^{o}$. The photoelectron intensity $I(k$,\textit{$\theta $},\textit{$\phi $}) was fitted by 
low-order polynomials to obtain the smooth background intensity 
\cite{Len:1999,Len:1997}:

\begin{equation}
\label{eq4}
I_{0} \left( {k,\theta } \right) = \left( {a_{0} + a_{1} k + a_{2} k^2} 
\right)\left( {b_{0} + b_{1} \cos \theta + b_{2} \cos 3\theta } \right).
\end{equation}

Three kinds of $\chi $ were obtained from this fitting: $\chi _{A} $ by 
fitting the second factor of Eq. (4) to a scanned-angle pattern $I_{k} 
\left( {\theta ,\phi } \right)$ at each fixed $k$ \cite{Terminello:1993}, 
$\chi _{B}$ by fitting the first factor to a scanned-energy curve $I_{{\rm 
{\bf \hat {k}}}} \left( k \right)$ at each fixed direction ${\rm {\bf \hat 
{k}}}$ \cite{Tong:1995} and $\chi _{C} $ by fitting both factors to 
the full data set of $I\left( {k,\theta ,\phi } \right)$ at one time, with 
the last expected to be the most accurate from an \textit{a priori} point of view 
\cite{Len:1999}. The $k$-differences from $\chi _{C} $ were also used 
for DPH in what we will term Method D (i.e., $\chi _{D} = \delta \chi _{C} 
)$. Since low-frequency fringes due to FS events in $I_{{\rm {\bf \hat 
{k}}}} \left( k \right)$ are automatically removed in Method B 
\cite{Wei:1994}, the resulting $I_{0}$ inherently deviates from the 
true $I_{0}$ defined as the intensity without scattering, especially in the 
FS direction. In addition, since $I_{k} \left( {\theta ,\phi } \right)$ and 
$I_{{\rm {\bf \hat {k}}}} \left( k \right)$ are independently normalized 
without considering the continuity of $\chi $ in the whole sampled ${\rm 
{\bf k}}$ space in Methods A and B, they could degrade 
holographic fringes in $I_{{\rm {\bf \hat {k}}}} \left( k \right)$ and $I_{k} 
\left( {\theta ,\phi } \right)$, respectively. By contrast, Method C takes 
into account the continuity of $\chi $ over the whole data set, but the FS 
peaks remain in $\chi _{C} $; however, they should be eliminated in $\chi 
_{D}$. The original transform of Eq. (\ref{eq1}) was used for all four data 
sets; but to avoid the abrupt truncation of the integral in Eq. (\ref{eq1}), $W$ was 
taken to be the product of a Gaussian function of $k$ and a Hanning function 
$\cos ^2\theta $, with an additional multiplication by $r$ to make atoms at 
larger distances more visible.

\begin{figure}[htbp]
\centerline{\includegraphics[width=3.5in]{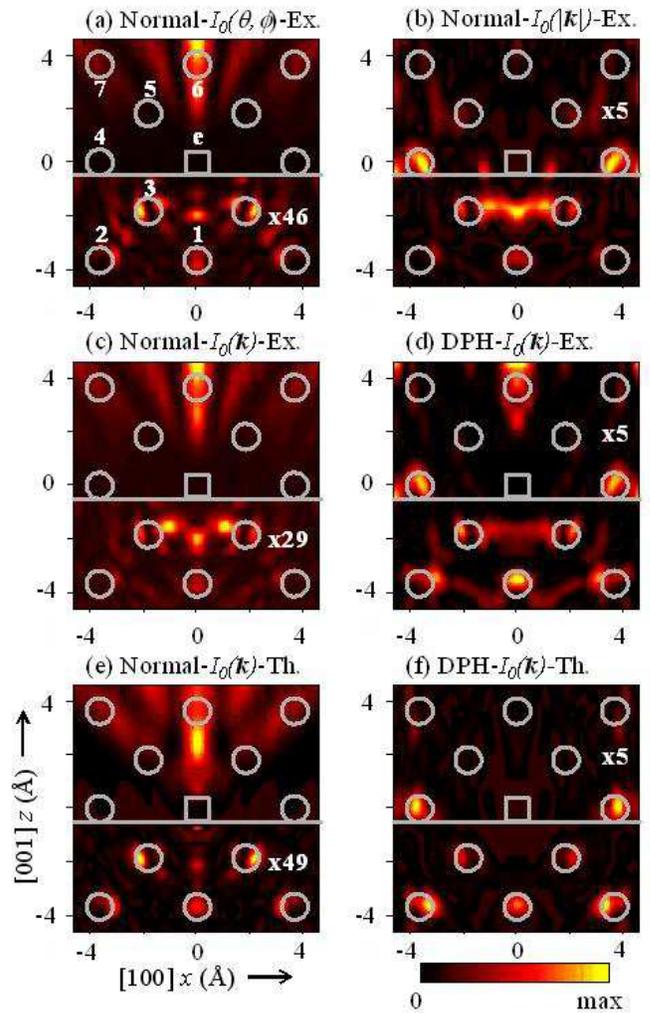}}
\caption{Atomic images in the vertical (100) plane of Cu (001) reconstructed from Cu 
3p holograms obtained by Methods A-D, as described in the text. The emitter 
and scatterer positions are indicated by squares and circles, respectively, 
and various near-neighbor atoms are numbered. Image intensities above or 
below $z_{c} ={\rm t}$0.5 {\AA} have been rescaled by the factor shown in 
each panel, with this factor being determined so as to make the maximum 
intensities above and below $z_{c} $ equal. Experimental images (a) Image 
from Method A: normal holography with $I_{0} $ determined by fitting the 
angular variation of Eq. (4) at each $k$ value. (b) Image from Method B: 
normal holography with $I_{0} $ determined by fitting the $k$ variation 
along each direction. (c) Image from Method C: normal holography with $I_{0} 
$ determined by fitting both the angular and k variation. (d) Image from 
Method D: differential holography, with $I_{0} $ as in (c). Theoretical 
images: (e) As (c) but theoretical.}
\label{fig2}
\end{figure}

Figure \ref{fig2} shows atomic images reconstructed 
from \textit{$\chi $}$_{{\rm A}}$-\textit{$\chi $}$_{D}$ in the vertical (100) plane of Cu(001). In Methods 
A and C, only elongated features related to FS effects from atoms of types 6 
and 7 are observed above $z_{c}=-0.5$ {\AA} (the arbitrary location 
of a change in image multiplication). This is consistent with a previous PH 
study of W(110) \cite{Len:1999}, in which Method C was used. By 
contrast, it has been reported \cite{Terminello:1993} that FS atoms of 
type 5 have been imaged via Method A from Cu 3p holograms for Cu(001) 
obtained at 9 energies. Even if possible differences in the two sets of 
experimental data are taken into account, it is difficult to conclude from 
our results that the images of these FS atoms can be resolved from strong 
artifacts via Method A. Below $z_{c}$, several peaks near the BS positions 
1-3 are observable for A and C among various strong artifacts, but only with 
the help of higher image amplifications of 46 and 29, respectively.

In Methods B and D, image intensities are stronger in the BS region, with 
the relative image amplification factors being reversed in sense and smaller 
at $\times 5$ compared to A and C. In Method D, a strong, somewhat elongated 
peak is observed at the FS position 6, with weaker features that appear to 
be associated with atoms 7 also present in the corners of the image. In both 
B and D, two strong peaks are observed at the SS positions 4 above $z_{c} $ 
and five peaks are observed at the BS positions 1-3 below $z_{c} $. However, 
the most intense features in Method B are the artifacts between the two 
nearest BS atoms of type 3. In Method D, by contrast, the five strongest 
peaks below $z_{c} $ are of roughly equal intensity and correspond 
reasonably well to the near-neighbor BS atoms. Therefore, we find Method D 
to be the most robust for imaging both SS and BS atoms (as well as to some 
degree also FS atoms 6), even if there are shifts in position of 
approximately 0.1 {\AA} for type 1, 0.6 {\AA} for 2, and 0.3 {\AA} for 3. 
Such peak shifts relative to the true atomic positions, as observed in all 
methods, can be attributed to the present neglect of corrections for both 
the scattering phase and the inner potential.

For comparison with experiment, we have also performed MS simulations of 
$I({\rm {\bf k}})$, using a cluster method fully described elsewhere 
\cite{Chen:1998}. The theoretical $I_{0}$ was obtained simply as the 
square of the zeroth-order wave function without scattering. Images 
reconstructed from the theoretical $\chi $ and $\delta \chi $ via Methods C 
and D are shown in Figs. \ref{fig2}(e) and (f) and 
can be compared with Figs. \ref{fig2}(c) and (d), 
respectively. The main features in Figs. \ref{fig2}(c) and (d) are well reproduced by our simulations, although the artifacts 
between the atoms 3 are much stronger in experiment for C, and the relative 
intensity in the region of FS atom 6 is stronger in experiment for D. Even 
though the ideal \textit{$\chi $} was used for image reconstruction, no atomically-resolved 
SS or FS peaks are observable in Fig. \ref{fig2}(e). 
Therefore, the corresponding artifacts in Fig. 
\ref{fig2}(c) are not purely due to the uncertainties 
in the experimental data and any errors in the $I_{0}$ subtraction, but must 
have their origin in the MS effects and basic imaging algorithm. On the 
other hand, Fig. \ref{fig2}(f) exhibits well-resolved 
peaks at the SS and BS positions. Since there are no artifacts below 
$z_{c}$ in Fig. \ref{fig2}(f), the artifacts in Fig. 
\ref{fig2}(d) are by contrast considered to be purely 
due to the experimental noise and other non-idealities in the data analysis.

We have also generated full three-dimensional atomic images from the 
experimental data via $\chi _{D}$, although length limitations prevent 
showing these here. In these images, we find in addition to the atoms 1-4 
and 6 in Fig. \ref{fig2}, two other types of 
near-neighbor BS and SS atoms located in the vertical (110) plane (denoted 
types 2' and 4' and situated in the same horizontal layers as 2 and 4, 
respectively). All of these atoms are reasonably well reconstructed, with 
only a few, such as 2, being significantly shifted in position, but most 
within a few tenths of an {\AA} of the correct positions in all directions. 
The overall positional errors for all of the atoms compared to the known Cu 
lattice can be summarized as (radial location shift in $xy)$/(vertical 
location shift in $z)$, and are: 0.0 {\AA}/0.1 {\AA} for atoms 1, 0.6 
{\AA}/0.1 {\AA} for 2, 0.3 {\AA}/0.1 {\AA} for 2', 0.2 {\AA}/0.1 {\AA} for 
3, 0.1 {\AA}/0.0 {\AA} for 4, 0.3 {\AA}/0.0 {\AA} for 4', and 0.0 {\AA}/0.4 
{\AA} for 6. As a further indication of the overall image quality obtained 
by DPH, the reader is referred to an animated comparison of 3D images for 
the four approaches of Figs. 2(a)-(d), in which DPH is alone in imaging 
approximately 15 near-neighbor atoms \cite{Animated:1}.

Finally, we compare DPH with a very recently introduced approach for PH 
termed near-node holography \cite{Wider:2001}, in which FS effects are 
suppressed by using a special experimental geometry with electron exit 
nearly perpendicular to light polarization. Although this technique is 
promising, DPH has the advantages that it does not require a special 
experimental geometry or s-subshell-like form for the photoelectric cross 
section, that it seems to yield images of as good or better quality 
\cite{Animated:1,Wider:2001} and that it can be used in 
other types of holography in which polarization cannot be varied.

In summary, we have demonstrated differential photoelectron holography (DPH) 
as a powerful method for overcoming the FS problem in PH and enhancing image 
quality for any kind of system in which FS can arise, as for example, bulk 
emission and buried interfaces. This method should also be helpful in other 
types of electron holography in which energy can be stepped in a controlled 
way (e.g. Kikuchi \cite{Wei:1994} or LEED \cite{Saldin:1998} 
holography). The reconstructed images for Cu 3p/Cu(001) demonstrate that DPH 
is successful in suppressing the FS effects so as to image SS, BS, and to 
some degree also FS, atoms with accuracies of 0.1-0.6 {\AA}.

This work was supported in part by the Director, Office of Energy Research, 
Basic Energy Science, Materials Sciences Division of the U. S. Department of 
Energy under Contract No. DE-AC03-76SF00098. S.O. and Y.N. also acknowledge 
the support of the Japan Society for the Promotion of Science (Grant No. 
JSPS-RFTF 98R14101).

\end{document}